\documentclass[graybox, envcountchap, authoryear, natbib]{svmult}

\usepackage{mathptmx}       % selects Times Roman as basic font
\usepackage{makeidx}         % allows index generation
\usepackage{amsmath}     % needed for \gtrsim symbol
\usepackage{amssymb}     % needed for \gtrsim symbol
\usepackage{graphicx}        % standard LaTeX graphics tool
\usepackage{multicol}        % used for the two-column index

\usepackage[bottom]{footmisc}% places footnotes at page bottom

\usepackage{natbib}

\makeindex             % used for the subject index
\newcommand{\aj}{AJ}
\newcommand{\apj}{Ap.J}
\newcommand{\apjl}{Ap.J Letters}
\newcommand{\apjs}{Ap.J Supplement}

\newcommand{\mnras}{MNRAS}
\newcommand{\aap}{A\& A}
\newcommand{\apss}{Ap \& SS}
\newcommand{\nat}{Nature}
\newcommand{\aapr}{A\&A~Rev.}      
\newcommand{\Msun}{M_{\odot}}
\newcommand{\HI}{H$\,$I}
\newcommand{\Rvir}{R_{\rm vir}}
\newcommand{\Vvir}{V_{\rm vir}}
\newcommand{\Mvir}{M_{\rm vir}}
\newcommand{\enzo}{\texttt{Enzo}}
\newcommand{\ramses}{\texttt{Ramses}}
\newcommand{\art}{\texttt{Art}}
\newcommand{\arepo}{\texttt{Arepo}}
\newcommand{\gizmo}{\texttt{Gizmo-PSPH}}
\newcommand{\gasoline}{\texttt{Gasoline}}

\begin{document}

\title*{Gas Accretion and Angular Momentum}
\author{Kyle R. Stewart}
\institute{Kyle R. Stewart \at Department of Mathematical Sciences, California Baptist University, 8432 Magnolia Ave., Riverside, CA 92504, USA, \email{kstewart@calbaptist.edu}}
\maketitle

\abstract{In this chapter, we review the role of gas accretion to the acquisition of angular momentum, both in galaxies and in their gaseous halos.
We begin by discussing angular momentum in dark matter halos, with a brief review of tidal torque theory and the importance of mergers, 
followed by a discussion of the canonical picture of galaxy formation within this framework, where halo gas is presumed to shock--heat to the virial 
temperature of the halo, following the same spin distribution as the dark matter halo before cooling to the center of the halo to form a galaxy there.  
In the context of recent observational evidence demonstrating the presence of high angular momentum gas in galaxy halos, 
we review recent cosmological hydrodynamic simulations that have begun to emphasize the role of ``cold flow'' accretion---anisotropic 
gas accretion along cosmic filaments that does not shock--heat before sinking to the central galaxy.  We discuss the 
implications of these simulations, reviewing a number of recent developments in the literature, and 
suggest a revision to the canonical model as it relates to the expected angular momentum content of gaseous halos around galaxies.
}

\section{Introduction}
\label{intro}
In the standard Lambda Cold Dark Matter (LCDM) paradigm, galaxies form at the center of extended dark matter halo.  These halos grow hierarchically through
halo mergers (including embedded galaxies) as well as by diffuse accretion of dark matter and gas from the cosmic web.  Diffuse infalling gas is expected to 
shock--heat to the virial temperature of the halo, mixing within the halo until it virializes, with gas eventually cooling out of this hot gaseous halo, 
sinking to the center of the halo and onto the central galaxy \citep[e.g.][]{Binney77,ReesOstriker77,Silk77,WhiteRees78,WhiteFrenk91,MallerBullock04}.
Under this (admittedly simplified) picture of galaxy formation, it is expected that the inflowing gas (and thus the virialized hot gaseous halo) should share the same angular momentum distribution 
as the inflowing dark matter.  Conserving angular momentum, the galaxy that ultimately forms should also have specific angular momentum similar to that of the dark matter halo, resulting in a rotationally supported 
disk galaxy (in many cases) with a spin proportional to the dark matter halo \citep{Mestel63,FallEfst80,MoMaoWhite98}, the statistical properties of which have been well--studied via dissipationless cosmological
$N$-body simulations \citep[e.g.][]{Bullock01, Vitvitska02, Maller02,Avila-Reese05,DOnghia07,Bett10,Munoz-Cuartas11,Ishiyama13,Trowland13,Kim(Choi)15,ZjupaSpringel16}.

However, in recent years, advances in galaxy formation theory (both in analytic work and via cosmological hydrodynamic simulations) have begun to emphasize the importance 
of the filamentary nature of gas accretion onto massive galaxies, particularly at high redshift when cosmic filaments are significantly narrower and denser than in the local universe.  
Filamentary gas accretion, though diffuse, may be dense enough to allow cold streams to maintain cooling times shorter than the compression time to establish a stable shock, leading 
to what has been referred to as ``cold flows'' or ``cold mode'' gas accretion that can quickly penetrate from the virial radius of a dark matter halo all the way to the inner galactic region of the halo
\cite[e.g.][]{Keres05, DekelBirnboim06, Ocvirk08, Brooks09, Dekel09, FGKeres10, FG11, vandeVoort11,Hobbs15,vandeVoort15}.  

While there has been some contention in recent years
as to whether or not these cold streams are truly capable of delivering unshocked gas directly onto the galaxy, without heating in the inner regions of the halo \citep[e.g.,][]{Torrey12,Nelson13,Nelson16},
the importance of distinguishing between these dense filamentary forms of gas accretion to galaxy halos (verses isotropic ``hot mode'' gas accretion) 
remains a crucial one for understanding galaxy formation.
In particular---as it relates to this chapter---under this developing paradigm of filamentary versus isotropic gas accretion, 
halo gas (and particularly gas accreted in the cold mode) tends to show considerably higher specific angular momentum than
the dark matter in the halo 
\citep{Chen03, SharmaSteinmetz05,Keres09,KeresHernquist09,Agertz09,Brook11, Stewart11b, Kimm11,Stewart13,Codis15,Danovich15,Prieto15,Teklu15,Tillson15,Stewart16}.  
In this picture, the resulting angular momentum of simulated stellar disks may be significantly different than that of the accreted gas, in part because feedback effects preferentially expel 
low angular momentum gas from galaxies \citep[e.g.][]{MallerDekel02,Governato10,Brook11,Eris}, such that the total cumulative spin of a growing galactic disk may not be expected 
to match the cumulative spin of accreted dark matter or gas to the virial radius of the halo.  

As a result, this emerging picture of galaxy grown seems to be in tension with the canonical picture in which the spin of the accreted gas (and ultimately, the galaxy) mimics the dark matter.  
Thus, a modified picture of angular momentum acquisition that takes into account these distinctions between filamentary and isotropic gas accretion 
seems to be developing in the literature.  While this picture is by no means fully developed, nor universally agreed upon, one of the main goals of this review is to serve as a 
synthesis for a number of recent theoretical works (primarily utilizing the results from hydrodynamic cosmological simulations) in order to present a consistent new picture for angular momentum acquisition to 
galaxies, as well as the expected angular momentum content of gaseous halos around galaxies.

The outline of this review is as follows.  We will begin in section \ref{DM} by briefly reviewing the origin of angular momentum 
in dark matter halos within the framework of Lambda Cold Dark Matter cosmology (LCDM), including a discussion of tidal torque theory (TTT), 
the role of mergers, and studies of dark matter halos from cosmological dissipationless $N$-body simulations.  
In section \ref{galaxies} we will review the canonical model for galaxy formation
in LCDM, which builds upon these properties of dark matter halos as a means of understanding the process by which 
gas within dark matter halos shock--heats, dissipates energy, and ultimately sinks to the center of the halo's gravitational potential to form stars (galaxies) there.
We will also discuss some of the historic challenges in simulating galaxies with realistic angular momentum in hydrodynamic cosmological simulations.
The most in-depth portion of this review is section \ref{CGM}, which begins by briefly discussing some of the observational challenges to this canonical picture---namely,
a number of recent observations of coherent co-rotation and high angular momentum gas in galaxy halos.  This is followed by a deeper discussion of recent
studies of hydrodynamic cosmological simulations that have begun to demonstrate a 
need to update the canonical picture of gas accretion onto galaxies (in large part by emphasizing the importance of ``cold flow'' filamentary gas accretion) and how 
these modifications are in better alignment with recent
observations.  We summarize and conclude in section \ref{conclusion}.

\section{Angular Momentum of Dark Matter Halos}
\label{DM}
Before address the complications of gas dynamics on the angular momentum acquisition of galaxies, let us review the 
theoretical picture for the origin of angular momentum in dark matter halos in which galaxies are embedded.  We will begin by briefly discussing 
some the general characteristics of the spin of dark matter halos as derived from cosmological dissipationless $N$--body simulations, before 
reviewing the tidal torque model in \S\ref{DM_TTT}, and the role of major and minor mergers in angular momentum acquisition in 
\S\ref{DM_mergers}.  

Within the framework of LCDM, the universe forms hierarchically, with less massive dark matter halos forming first, and halos 
merging together to form more massive halos over time \citep[e.g.,][]{Peebles82,Blumenthal84,Davis85}. 
In this paradigm, since dark matter dominates the matter density of the universe, 
galaxies are expected to reside within the centers of massive dark matter halos.  Thus, an important starting point for understanding the 
angular momentum of galaxies is to determine the angular momentum of dark matter halos in which galaxies reside, typically characterized by 
the dimensionless spin parameter \citep{Peebles69},
\begin{equation}
\lambda_P\equiv \frac{J |E|^{1/2} }{GM^{5/2}}
\label{eq:spin_Peebles}
\end{equation}
or by the revised version of the spin parameter, first presented by \cite{Bullock01}, 
rewritten in terms of a dark matter halo's radius and virial velocity\footnote{The revised spin parameter from \cite{Bullock01} was first 
introduced as $\lambda'$, but for the purposes of our discussions in this chapter, we will drop the prime and adopt this revised spin parameter as simply $\lambda$.}:
\begin{equation}
\label{eq:spin}
\lambda = \frac{j}{\sqrt{2} \Vvir \Rvir}
\end{equation}
where $j=J/M$ is the magnitude of the specific angular momentum of the halo, $\Vvir=\sqrt{G\Mvir/\Rvir}$ is the circular velocity at the halo virial radius, 
and $\Rvir$ and $\Mvir$ are the halo virial radius and virial mass, respectively.

%>>>>>>>>>>>>>>>>>>>>>>>>>>>>>>>>>>>>>>>><<<<<<<<<<<<<<<<<<<<<<<<<<<<<<<<<
\begin{figure}[t]
\sidecaption
\includegraphics[width=0.6\textwidth]{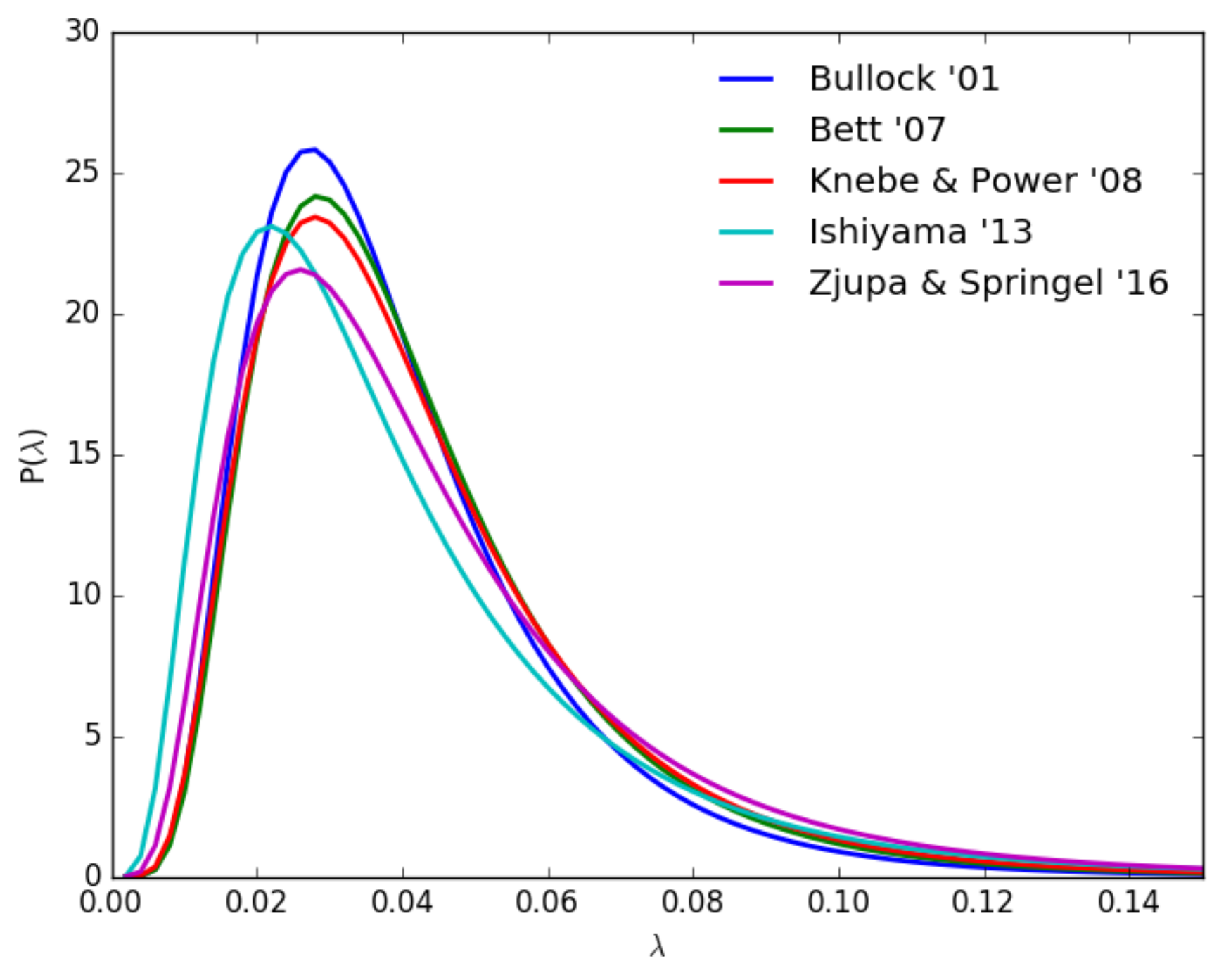}
\caption{Distributions of dark matter halo spin parameters from \cite{Bullock01,Bett07,KnebePower08,Ishiyama13} and \cite{ZjupaSpringel16}.
Shown here are the  best fits to a the log--normal function defined in Equation \ref{eq:lognormalspin}.  
Quantitative details vary among of the simulation (based on, e.g., variations in the definition of the dark matter halo virial radius). 
However, there is good agreement among simulations, with best--fit parameters $\lambda_0\simeq0.035$ and $\sigma\simeq0.5$.}
\label{fig_DMspin}       % Give a unique label
\end{figure}
%>>>>>>>>>>>>>>>>>>>>>>>>>>>>>>>>>>>>>>>><<<<<<<<<<<<<<<<<<<<<<<<<<<<<<<<<

Remarkably, a long history of studies of $N$-body simulations reveal that the spin parameters of dark matter halos do not show substantial trends with  
halo mass, redshift, or environment, 
\citep[e.g.][]{FallEfst80,BarnesEf87,Bullock01,Vitvitska02,Bett07, Maccio07, Berta08, Bett10}, which is a 
natural result of the expectations of Tidal Torque Theory as the origin of dark halo angular momentum (see \S\ref{DM_TTT}, below).
Instead, the distribution of halo spins from a number of simulations demonstrate a relatively good fit to a log--normal distribution:
\begin{equation}
P(\lambda) = \frac{1}{\lambda\sigma\sqrt{2\pi}} \exp{ \Big( -\frac{\ln^2{(\lambda/\lambda_0)}}{2\sigma^2} \Big)  }
\label{eq:lognormalspin}
\end{equation}
 with the distribution peaking at $\lambda_0\simeq 0.035$ with a Gaussian width of $\sigma\simeq0.5$.  For comparison, examples of best--fit parameters to 
this lognormal distribution from \cite{Bullock01,Bett07,KnebePower08,Ishiyama13} and \cite{ZjupaSpringel16}\footnote{Since \cite{ZjupaSpringel16} does not 
provide fitting parameters to the modified $\lambda'$ from \cite{Bullock01}, we compare here with their SO--halo lognormal fit to $\lambda_P$ from 
\cite{Peebles69}, given in Equation \ref{eq:spin_Peebles}, which appears qualitatively similar to their results for the Bullock definition, based on their Figure 9}, 
across a wide range in dark matter halo masses,
are shown in Figure \ref{fig_DMspin}, showing remarkable agreement among simulations. 
The mass distributions of angular momentum within dark matter halos has also been shown to fit a universal two--parameter angular momentum 
profile \citep{Bullock01}, based on the halo spin parameter as well as a halo shape parameter, though the underlying reason dark matter halos should
be fit to a nearly universal angular momentum distribution profile is still not well understood.

\subsection{Tidal Torque Theory}
\label{DM_TTT}
The foundational picture of the origin of angular momentum in large scale structures (such as galaxies and dark matter halo)s is provided by Tidal Torque Theory 
\citep[TTT, e.g.,][]{Hoyle49, Peebles69, Sciama55, Doroshkevich70, White84,BarnesEf87}.
In TTT \citep[for a more thorough recent review, see][]{Schafer09}, until the initial density perturbations in the universe
reach maximum expansion (turnaround), their angular momentum grows linearly with cosmic time as a consequence of the torques 
exerted by the tidal gravitational fields from neighboring overdensities.  After turnaround, structures decouple from the Hubble Flow and these overdensities 
collapse and virialize, conserving angular momentum in the process,
such that the angular momentum of individual halos may be predicted (within an order of magnitude) 
by the initial large--scale gravitational tidal torques before maximum expansion.  However, detailed analysis requires the inclusion of 
non--linear effects after turnaround, and thus is typically carried out by cosmological $N$--body simulations.

Under TTT, the total angular momentum acquired by a halo of a given
mass at turnaround is expected to scale as $L\propto M^{5/3}$ \citep{Peebles69}, 
with more massive halos acquiring more angular momentum, in part because it takes a long time for more massive halos to reach turnaround and 
decouple from the Hubble expansion.  Taken in combination with the
definition of the halo spin parameter in Equation \ref{eq:spin}, in
which $\lambda\propto J/MVR$,
as well as the virial scaling relations $V\propto (M/R)^{1/2}$ and
$R\propto M^{1/3}$,  one straightforward expectation from TTT is that the halo spin parameter should be independent of halo mass (and redshift, as long as the 
redshift in question is after turnaround).  This expectation of TTT has been well tested in a number of numerical studies using cosmological $N$-body simulations
of large scale structure, and is foundational to numerous semi--analytic models of galaxy formation
(see \S\ref{SAMs}), however, the importance of non--linear effects also lead to a number of quantitative disagreement between TTT and $N$-body simulations; 
for example, linear spatial correlations between the spins of halos on scales greater than about $1$ Mpc are
over-predicted in TTT by $\sim50-70\%$ compared to simulations \citep[e.g.,][]{Porciani02a,Porciani02b}.

More recently, \cite{Codis12} and \cite{Codis15} 
revisited the framework of TTT, emphasizing the importance of the anisotropic geometry of walls and filaments in the cosmic web.  
They found that the alignment direction of angular momentum in dark matter halos is dependent on the geometry of the cosmic web.  
The misalignments of accretion flows in the walls that collapse into filaments 
result in spin directions aligned \emph{with} filaments for low mass halos.  However, higher mass halos are strongly influenced by accretion (including mergers)
that flow along these filaments, resulting in spin directions that are \emph{perpendicular} to the filament.  (This will be an important theoretical 
framework to keep in mind as we discuss revisions to the canonical picture of angular momentum acquisition onto galaxies in \S\ref{CGM}.)

\subsection{Angular Momentum Acquisition via Mergers}
\label{DM_mergers}
An alternate way to study angular momentum acquisition in galaxy halos is by considering the impact of major and minor mergers during the 
hierarchical merger history of a given dark matter halo.  In this model, the final angular momentum of a dark matter halo
is determined by the sum of the orbital angular momentum of all merging satellites over the course of its accretion history,
and was found to match the spin distribution of dark matter halos from cosmological $N$-body simulations \citep{Vitvitska02,Maller02}.

One useful way to conceptualize this approach in a complementary role
to TTT (rather than a ``competing'' mechanism) is to consider
that the initial large scale tidal torques serve as a means of establishing the tangential velocities of infalling satellite halos.  Thus, rather 
than attempting to fully understand the role of tidal torques in setting the ultimate angular momentum of a dark matter halo, one may instead
focus on the particular kinematics of infalling satellites over the accretion history of the halo, which are in turn determined by the gravitational 
torques from the initial tidal field.

Under the merger model, galaxy halos tend to show larger variation in spin parameter than predicted in TTT.  Larger mergers contribute substantial 
angular momentum based on their orbital motion, leading to significant spikes in the spin parameter of a post--merger halo, typically followed by 
a steady decline in spin parameter during epochs of gradual smooth accretion.  Since major mergers can lead to a significant redistribution of angular momentum,
under this model the high end of the angular momentum distribution of a dark matter halo is largely determined by the orbital angular momentum 
of its last major merger, with minor mergers (presumably from random infall directions) tending to contribute to the low end of the distribution.

\section{The Angular Momentum of Galaxies}
\label{galaxies}
In the canonical picture of galaxy formation, the initial distribution of baryons in the universe matched that of the dark matter, such that when 
dark matter overdensities collapse and virialize, gas is also affected by the same large--scale tidal fields as the dark matter, resulting in a similar distribution
of baryonic matter as that of the dark matter.  However, since the halo gas is dissipational (unlike dark matter), it is capable of radiating away orbital and thermal energy,
sinking to the center of the halo's gravitational potential, until it is ultimately cold and dense enough to form stars.  Thus, all galaxies are thought to be 
embedded at the center of a massive dark matter halo, with the sizes, luminosities, morphologies and angular momentum content of those galaxies 
owing to the details of their formation---which are likely to be correlated in some way with the formation of the dark matter halo.
In this section, we will detail the canonical model for how this link between dark matter halo formation and galaxy formation is thought to operate, including
the importance of this model in laying the foundation for semi-analytic models (SAMs) of galaxy formation, and discussing the levels of agreement between
such models and observations.  
We will then give a brief review of the challenges and achievements in attempting to simulate galaxy formation directly with hydrodynamic cosmological simulations.

\subsection{Modeling Gas Accretion onto Galaxies}
\label{SAMs}
The classic picture of galaxy formation \citep[e.g.,][]{FallEfst80, WhiteFrenk91, MoMaoWhite98} attempts to model the formation of galactic disks inside the hierarchical 
framework of LCDM by making a few assumptions about the relationship between the baryons and dark matter.
In these relatively simple models it is possible to reproduce a number of observable properties of spiral galaxies (e.g., the slope and scatter of the Tully--Fisher relation) 
as well as damped Ly$\alpha$ absorbers, while making as few underlying assumptions as possible.  For example, \cite{MoMaoWhite98} use the following fundamental assumptions: 

\begin{enumerate}
\item As the halo forms, the gas initially relaxes into an isothermal distribution.  Further gas accretion is shocked to the virial temperature of the halo.  
Virialized gas subsequently cools, conserving angular momentum.
\item The specific angular momenta of galaxy disks are thus similar to their parent halos, $j_d\simeq j$ (alternatively, $\lambda_d\simeq\lambda$).  As a result, the total 
angular momenta of disks is expected to be a fixed fraction of that of the halo: $J_d/J \simeq M_d/M$.
\item Galaxy disks have masses that are a fixed fraction of roughly a few percent of the mass of their parent halos: $M_d/M \le 0.05$ % \lesssim -> \le
\item The resulting disk is assumed to be rotationally supported with an exponential surface density profile and $R_d\simeq \lambda\Rvir$.
\end{enumerate}

Building on this approach, more recent semi-analytic models include additional physical models such as 
supernova feedback that expels gas from galaxies, black hole growth and feedback that heats gas in galaxy clusters, estimation of 
the cooling radius and cooling rate out of the hot halo, as well as effects of galaxy mergers such as starbursts and morphological transformation 
\citep[recently, e.g.,][]{Cattaneo06, Croton06, Somerville08, Dutton12, Somerville12}.  By tuning the input parameters of these models on certain observational constraints \citep[e.g. tuning 
the chemical yield of supernovae to reproduces metallicities of stars in galaxies in][]{Somerville08}, it is possible to produce modeled galaxy populations that reproduce a great number 
of physical properties of galaxies: e.g., cold gas fractions, stellar ages, specific star formation rates, stellar mass functions, etc. 

While it is beyond the scope of this chapter to provide a more comprehensive review of semi-analytic models of galaxy formation, 
one important point for consideration (especially for our discussion in \S\ref{CGM}) is that while some more recent models do implement a 
distinction between ``cold mode'' accretion that 
does not shock--heat to the virial temperature, versus ``hot mode'' gas accretion which does shock--heat, the 
angular momentum of the halo gas (regardless of which ``mode'' is used)
is still modeled by $\lambda_d = \lambda_{\rm gas} = \lambda_{\rm DM}$.  
However, hydrodynamic simulations suggest that galactic outflows may preferentially expel low angular momentum gas from the centers of galaxies, keeping galaxy formation inefficient and stopping 
forming galaxies from universally creating massive bulges at early times.  In this case, even if galaxies initially form with $\lambda_d=\lambda_{\rm DM}$ at early times, 
this similarity would be expected to break over cosmic time, as outflows continue to preferentially remove low angular momentum gas from the galaxy (without similarly removing dark matter from the halo). 
Furthermore, while one might expect the overall spin of halo gas and dark matter to be in rough agreement,  
we will see in \S\ref{CGM_theory} that the detailed accretion geometry of different modes of of gas accretion (particularly the contribution of dense filamentary ``cold mode'' gas) results in a scenario
where $\lambda_{\rm gas}\ne\lambda_{\rm DM}$.

\subsection{Hydrodynamic Simulations of Galaxy Formation}
Early work in cosmological hydrodynamic simulations showed great difficulty in successfully simulating disk dominated galaxies.  In what is often referred to as the 
``angular momentum catastrophe'', simulations produced either spherical galaxies or disks with significantly lower angular momentum than the halo, with 
orbital angular momentum being transferred to the dark matter by dynamical friction before the baryons reach the center of the halo 
\citep[e.g.][]{Katz92, NavarroWhite94, Sommer-Larsen99, Steinmetz99, NavarroSteinmetz00, DOnghia06}.  Not surprisingly, these simulated galaxies also 
produced unrealistic rotation curves and failed to match other observational constraints, such as the Tulley--Fisher relation.

The alleviation of this problem seemed to be the inclusion of efficient star formation feedback, which preferentially removes low angular momentum gas 
(that would otherwise form stars) from the centers of galaxies during the formation process \citep[e.g.][]{Governato07, Scannapieco08, Brook11, Eris, Governato10,Ubler14,Christensen16}.
This feedback makes galaxies considerably less efficient at forming stars, also keeping them gas--rich for longer.  This, in turn, also helps alleviate the tension between the observed abundance of disk dominated galaxies
\citep[e.g.,][]{Weinmann06} with the frequency of major mergers derived from $N$-body simulations \citep[e.g.,][]{Stewart08,Fakhouri10}, as both direct hydrodynamic simulation as well as semi--empirical galaxy formation models
suggest that gas--rich major mergers may help \emph{build} angular momentum supported disks from the surviving merger remnant, rather than transforming pre-existing disks into spheroids 
\cite[][]{Robertson06a, Stewart09b, Hopkins09, Governato09}.
With these advances in star formation and feedback prescriptions (as well as more advance computational power), recent simulations have essentially eliminated the early angular momentum problem,
allowing hydrodynamic simulations to successfully produce bulgeless exponential disk galaxies with properties quite similar to those observed in the real universe \citep[e.g.,][]{Governato10, Brook11, Eris}. 

Most importantly for our discussion of angular momentum acquisition in galaxies (and their halos), 
recent hydrodynamic simulations have also begun to place growing emphasis 
on the different ``modes'' of gas accretion onto galaxies, especially at high redshift.
In what is labeled ``hot--mode'' accretion, gas continues to behave in the manner previously described, shock--heating to the virial temperature of the halo, mixing, and eventually cooling on the galaxy.
However, the main mode of gas accretion for most galaxies is thought to be via ``cold--mode'' (or ``cold flow'') accretion---where the inflowing gas streams originating from filamentary accretion are 
dense enough at high redshift to have cooling times shorter than the shocking compression timescales, resulting 
in direct gas accretion from the cosmic web, through the galaxy halo, and onto the outskirts of the galaxy \citep[e.g.,][]{Keres05, DekelBirnboim06, Brooks09}.
As a result, this cold mode gas does not necessarily  mix with the existing gaseous halo, and so the specific angular momentum of gas that accretes onto the central galactic disk might \emph{not} 
be well matched by that of the dark matter halo, as previously assumed.  
We will discuss possible implications of this dual mode of accretion for understanding angular momentum in galaxy halos in \S\ref{CGM_theory}.

\section{Angular Momentum of Gaseous Halos}
\label{CGM}
The previously described model for galaxy formation (and angular momentum acquisition in particular) in \S\ref{SAMs} assumes that 
%cosmological gas accretion to galaxy halos 
%typically shock--heats to the virial temperature of the halo, allowing mixing in the halo such that the 
gaseous halos of galaxies should maintain a similar distribution of spin parameters to that of the dark matter (since the baryons and the dark matter both share the 
same initial tidal torques as an origin of their angular momentum).  Thus, if one were to consider the spin parameter of the gas in a galaxy halo 
%$\lambda_{\rm gas} = j_{\rm gas} / \sqrt{2} \Vvir \Rvir$, it should share
it should be the same distribution as the spin of the dark matter, which is well constrained from $N$-body simulations (Figure \ref{fig_DMspin}), such that
$\lambda_{\rm gas} \simeq \lambda_{\rm DM}$.
%Thus, the canonical model of galaxy formation assumes that the gas embedded in a dark matter halo represents a 
%fixed fraction of the mass and angular momentum of the dark matter, eventually cooling to form a galactic disk, such that 
%$M_{\rm d}/M = J_{\rm d}/J \simeq \lambda \simeq 0.035$. 
This simple theoretical picture has provided reasonable agreement between 
theory and observations, in terms of matching distributions of galaxy sizes and luminosities, with 
characteristic galaxy sizes expected to be $R_{\rm d} \sim \lambda\Rvir$ 
\citep[corresponding to $\sim10$ kpc for a galaxy halo with $\Rvir\sim300$ kpc, e.g.][]{FallEfst80,Bullock01,DuttonvandenBosch09}.

Having also reached the point where cosmological hydrodynamic simulations of galaxy formation (with properly tuned star formation and stellar feedback physics implemented) 
are able to produce galaxies with realistic disk scale lengths, bulge--to--disk ratios and overall angular momentum content---resolving the angular momentum 
catastrophe---one might think that our picture of angular momentum acquisition is reasonably complete.  However, recent observations have provided 
further complication to the issue of angular momentum acquisition, with numerous detections of baryons in galaxy halos 
with significantly \emph{higher} spin than either the galaxy or expectations for dark matter halos.
These frequent detections of high--spin baryons in galaxy halos would seem 
%Below, we briefly review several categories of 
%observations (though certainly not an exhaustive list) that demonstrates the high spin nature of baryons in galaxy halos 
%(compared to the stellar content of the galaxies, or the typical spin of dark matter halos), which would be 
difficult to explain under the assumption that $\lambda_{\rm gas} \simeq \lambda_{\rm DM}$.
%,before moving on to recent results from simulations that help explain these phenomenon.

In this section, we will discuss the angular momentum content not of the dark matter halo, nor the baryons in the stellar content of the galaxy, 
but instead the baryons present in the gaseous halo of the galaxy---i.e.~the circumgalactic medium (CGM).  We begin in
\S\ref{CGM_observations} by presenting the observational evidence for high angular momentum material in galaxy halos, challenging 
the classical picture of galaxy formation described above.  We then  
discuss recent advances in our understanding of galaxy formation theory that may help explain these observations in \S\ref{CGM_theory}.

%\subsection{Complicating the Issue: Observations of High Angular Momentum Gas in Galaxy Halos}
\subsection{Observations of High Angular Momentum Gas}
\label{CGM_observations}

%\subsubsection{Extended XUV and HI Disks in the Local Universe}
%{\bf PLACEHOLDER TEXT} 
In the local universe, some of these high angular momentum observations include
detections of extended \HI$ $ disks, XUV disks, and giant low surface
brightness galaxies \citep[e.g.,][]{Bothun87,Matthews01, Oosterloo07, 
ChristleinZaritsky08,Sancisi08,Lemonias11,Heald11,Holwerda12,Hagen16}, as well as low metallicity high angular momentum gas 
(presumably from fresh accretion) in polar ring galaxies \citep{Spavone10}.
For example, observations of UGC 2082 from \cite{Heald11} show a stellar disk diameter of $D_{25}=24$ kpc 
(defined by a surface brightness of 25 mag/arcsec$^2$) but  
the \HI$ $ disk (down to a minimum column density of $10^{20}$ cm$^{-2}$) has significantly higher specific angular 
momentum---being larger by roughly a factor of $\sim2$, $D_{HI} = 44$ kpc.  In an even more extreme example, \cite{Oosterloo07}
detected \HI$ $ disks as large as $\sim200$ kpc in diameter around
early type galaxies.
As another example, the giant lower surface brightness
galaxy UGC 1382 contains a low surface brightness stellar disk with a
$\sim38$ kpc radius, embedded in a $\sim110$ kpc \HI$ $ disk, residing
in a $\sim2\times10^{12}\Msun$ halo \citep{Heald11}. 
The high spin of such extended disk components is 
is somewhat difficult to understand within the context of the canonical model, in which $j_d\simeq j_{\rm gas}\simeq j_{\rm DM}$. 
Furthermore, \cite{Courtois15} also indicated that local extended \HI$ $ disks may be 
dependent on the galaxy's filamentary environment, suggesting there may be a fundamental distinction between 
the angular momentum content of filamentary accretion versus isotropic accretion.

%\subsubsection{Absorption Line Observations: $0.5<z<1.5$}
%{\bf PLACEHOLDER TEXT} 
At moderate redshift ($z\sim0.5$--$1.5$) there are a growing number of absorption line studies of the circumgalactic medium of 
galaxies that have begun to emphasize the bi--modal properties of absorbers 
\citep{Kacprzak10,Kacprzak12a,Kacprzak12b, Bouche12,Bouche13,Crighton13,Nielsen15,Diamond-Stanic15,Bouche16,Bowen16}, where
absorbers along a galaxy's major axis 
tends to show higher angular momentum inflows that are roughly co--rotating with the galactic disk and absorbers along a galaxy's minor axis 
tend to instead show observational signatures of outflowing gas.  Increasingly, a number of absorption system observations 
seem to be in agreement with models that include massive, extended structures with inflowing disk--like kinematics \citep[e.g.,][]{Bouche16,Bowen16}
%\subsubsection{Ly--$\alpha$ Blobs and Protogalactic Disks: $z>2$}
%{\bf PLACEHOLDER TEXT} 

At higher redshift ($z\sim2$--$3$) kinematic studies of Ly$\alpha$ ``blobs'' have observed large scale rotation that seems consistent 
with high angular momentum cold gas accretion \citep{Martin14,Prescott15}, and there have also been recent detections of 
massive protogalactic gaseous disks that are kinematically linked to gas inflow along a cosmic filaments \citep{Martin15,Martin16}.
(We will show in \S\ref{coldflowdisks} how these observations are strikingly similar to theoretical expectations for inspiraling cold streams from cosmological simulations.)  

Taken together (though this is by no means an exhaustive list), such observations show growing evidence for the existence of coherent rotation with high angular momentum for 
cold halo gas, in stark contrast to the theoretical picture where halo gas should have specific angular momentum similar to that of the galaxy and the dark matter halo.

%\subsection{Revising The Model for Gas Accretion in Galaxy Halos: ``Cold Flow'' Gas Accretion and Angular Momentum}
\subsection{``Cold Flow'' Gas Accretion and Angular Momentum} %edited for brevity
\label{CGM_theory}
Recent advances in galaxy formation theory and cosmological simulations have also begun to complicate the picture of galaxy formation presented
in \S\ref{galaxies}, with growing emphasis on multiple modes of accretion to galaxy halos.  
While isotropic ``hot--mode'' accretion continues to behave in the manner previously assumed, 
it has also been shown that anisotropic ``cold--mode'' accretion along cosmic filaments may have cooling times shorter than the compression timescales 
for creating a stable shock \cite[e.g.][]{Binney77,Keres05, DekelBirnboim06, Ocvirk08, Brooks09, Dekel09, FGKeres10, FG11, Stewart11a, vandeVoort11,Hobbs15,vandeVoort15}.  
As a result, this cold filamentary accretion does not spend sufficient time in the halo to become well mixed before sinking towards the central galaxy\footnote{Insofar as the angular momentum of spheroids at low redshift are 
still thought to be most strongly correlated with the merger history of its dark matter halo,  we note that
the following discussion mostly pertains to the newfound importance of filamentary cold accretion to the growth of massive disk--dominated galaxies,
or to the properties of gaseous halos of galaxies.}.

%\subsubsection{Theoretical Motivation: Angular Momentum and Filamentary ``Cold Flow'' Gas Accrection}
\label{coldflows}
In one of the seminal papers outlining the importance of ``cold mode'' gas accretion to galaxies, \cite{Keres05} suggested that the 
angular momentum of filamentary cold gas accretion may be substantially different than that of isotropic ``hot mode'' accretion, however for a more 
thorough investigation of the angular momentum of these different types of gas accretion, we must look to more recent results (in part owing to the 
need for superior numerical resolution in simulations before an analysis of 
angular momentum in gas accretion could be considered reasonably robust).  

%>>>>>>>>>>>>>>>>>>>>>>>>>>>>>>>>>>>>>>>><<<<<<<<<<<<<<<<<<<<<<<<<<<<<<<<<
\begin{figure}[t]
\sidecaption
\includegraphics[width=1.0\textwidth]{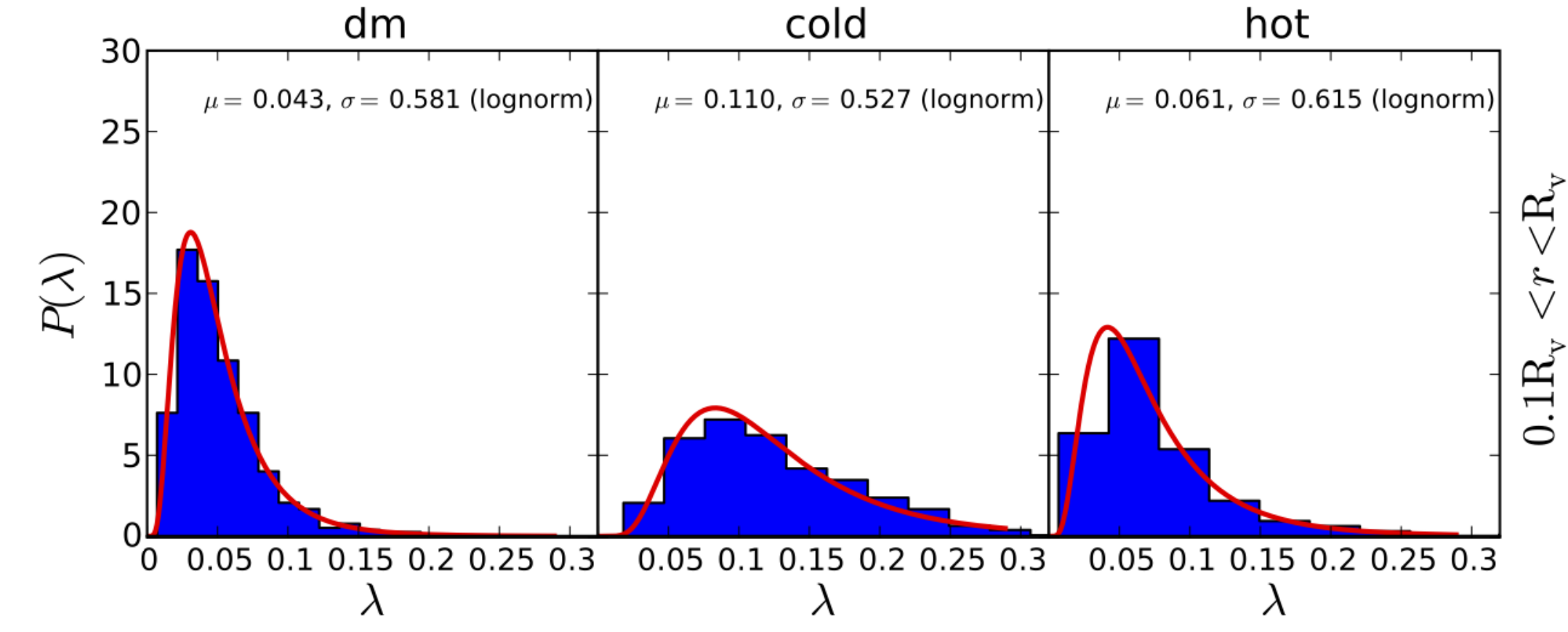}
\caption{Modified version of Figure 6 from \cite{Danovich15}, showing distributions of halo spin parameters (for material in the halo but not the central $0.1\Rvir$) 
in dark matter (left), cold gas (middle) and hot gas (right) from $29$ high resolution cosmological hydrodynamic ``zoom--in'' simulations between $z=4-1.5$.   
Red curves represent best fit lognormal fits to the data.  The mean spin parameter of each best fit curve ($\mu=\langle\lambda\rangle$) is given for each component, 
demonstrating that cold halo gas, $\langle\lambda_{\rm cold}\rangle=0.110$, has significantly higher specific angular momentum than the dark matter in the same region: $\langle\lambda_{\rm dm}\rangle=0.043$.
This fundamental result is in close agreement with a number of other studies \citep{Stewart11b,Pichon11,Kimm11,Stewart13,Stewart16}, which implement a 
variety of simulation codes and feedback implementations to study the spin of cold halo gas 
compared to that of the dark matter halo.
}
\label{fig_gasspin}       % Give a unique label
\end{figure}
%>>>>>>>>>>>>>>>>>>>>>>>>>>>>>>>>>>>>>>>><<<<<<<<<<<<<<<<<<<<<<<<<<<<<<<<<

\cite{Stewart11b,Stewart13} studied $4$ high--resolution cosmological hydrodynamic simulations of roughly Milky Way size halos run to $z=0$ 
\citep[using the smooth particle hydrodynamics (SPH) code \gasoline,][]{Gasoline}, with particular emphasis on the distinction between ``cold mode'' gas 
accretion (which is typically more filamentary and thus anisotropic)
versus ``hot--mode'' gas accretion (typically more isotropic).
%with an emphasis on observational signatures of cold mode gas accretion and 
They found that cold mode gas in galaxy halos 
contains significantly higher specific angular momentum than the dark matter, $\lambda_{\rm cold} \sim 4 \lambda_{\rm DM}$, 
and also has noticeably higher spin than the hot mode accretion, $\lambda_{\rm cold} \sim 2 \lambda_{\rm hot}$ (also see Figure \ref{fig_gasspin}).  
As a preliminary 
look at origin of this discrepancy, they compared the spin of recently accreted isotropic versus anisotropic dark matter and found a qualitatively similar distinction.

They also found that fresh accretion (both gas and dark matter) contained $\sim2$ times 
higher angular momentum than the entire halo, with $\lambda\sim0.1$ (rather than the canonical $\lambda\simeq0.04$ for the entire halo).
Furthermore, even at accretion to the virial radius, cold gas enters the halo with $\sim70\%$ more specific angular momentum than dark matter, and 
%by tracking the motion of gas through the halo, cold mode gas 
has a relatively short sinking time from the virial radius to the 
galactic disk (only $\sim1-2$ halo dynamical times).  They argue that this combination naturally explains the high angular momentum nature of cold halo gas, compared to dark matter:
dark matter in galactic halos represents a cumulative process of past accretion, while cold gas currently in a galactic halo both entered the halo with higher 
specific angular momentum than dark matter, and also represents recent accretion rather than a cumulative sum of all past accretion.   As a result of the coherent 
high nature of this cold inflow, they also reported the formation of transient ``cold flow disk'' structures in their simulations: massive extended planar structures of inflowing
cold halo gas (not rotationally supported) that are often warped with respect to the central galaxy---aligned instead with the 
angular momentum of the inflowing cold filamentary gas.

In a set of companion papers, \cite{Pichon11} and \cite{Kimm11} analyzed a statistical sample of $\sim15,000$ halos at $z>1.5$ (from a somewhat lower resolution simulation)
as well as $\sim900$ intermediate resolution halos and $2$ high resolution zoom--in simulations run to $z=0$ \citep[using the \ramses$ $ code,][]{Ramses},
finding that the ability of gas to radiative cool significantly alters the angular momentum transport of gas and dark matter into halos of a wide ranges of masses,
primarily due to the dense filamentary nature of cold gas accretion.  They reason that due to the asymmetry of cosmic voids, gas and dark matter 
flowing out of voids onto cosmic filaments gain a net transverse velocity, which acts as the seed of a halo's angular momentum as the material 
subsequently flows along the filament onto a nearby (gravitationally dominant) dark matter halo.  In this picture, material initially farther away from the filament
will gain a larger transverse velocity by the time it impacts the filament, naturally providing more angular momentum at later times, but with coherent 
direction (since the filament direction does not substantially change orientation over cosmic time).

While they find that both gas and dark matter tend to deliver similar specific angular momentum (as would be expected from TTT) the newly 
accreted material (at any given time) has significantly higher spin than that of the halo as a whole.  
Specifically, at high redshift, a large discrepancy  between
the spin of gas and dark matter in the halo arises, in agreement with previous work \citep{Stewart11b}, with $\lambda_{\rm gas} \sim 2-4 \lambda_{\rm DM}$, depending on halo mass. 
This discrepancy  is thought to be due to the coherent dense filamentary accretion of gas along cosmic filaments, without
shock--heating to redistribute the angular momentum through the entire halo.  Over time, both dark matter and baryons (including all gas and stars)
lose a significant amount of angular momentum, primarily via vector cancellation, as freshly accreted material is never perfectly aligned 
with that of the halo as a whole.

Using $2$ intermediate resolution simulations, \cite{Sales12} studied $100$ roughly Milky Way size halos run to $z=0$ (using the \ramses$ $ code).  They 
noted that the angular momentum of different gas components upon infall may not necessarily  be indicative of how these modes contribute to galaxy type.
They found that galaxy morphology was most strongly correlated with the \emph{coherent alignment} of angular momentum over cosmic time.  
Specifically, they found that most disk--dominated galaxies
formed their stars from hot--mode gas that shock--heated and eventually cooled onto the galaxy at later times, while more spheroids
were formed from cold--mode gas that sinks onto the galaxy more quickly, forming stars at a much earlier time; as a result, later episodes of accretion 
may not be very well aligned with the galaxy, leading to a spheroidal morphology.

\cite{Danovich15}, building upon previous work \citep{Danovich12}, analyzed $29$ zoom--in simulations at $z>1.5$ \citep[using the \art$ $ code,][]{Art1,Art2},
focusing on the angular momentum transport from the cosmic web onto massive galaxies, which they comprehensively detailed through four distinct phases, outlined below.
\begin{enumerate}
\item According to TTT, the spatial dependence of the angular momentum vector components, $J_i$ are given by the antisymmetric tensor product: 
$J_i \propto \epsilon_{ijk}T_{jl}I_{lk}$, where $T_{jl}$ is the tidal tensor and $I_{lk}$ is the inertial tensor.  In the principle coordinates of the tidal tensor, the 
angular momentum is thus proportional to the difference in the corresponding eigenvalues of the inertial tensor:
 $J_1 \propto T_{23}(I_3-I_2)$, 
 $J_2 \propto T_{13}(I_3-I_1)$, 
 $J_3 \propto T_{12}(I_2-I_1)$.  Under the assumption that the underlying tidal tensor should be approximately  the same for both dark matter and gas, any inherent differences 
in the specific angular momentum of dark matter versus gas (outside the virial radius, in the regime of TTT), should result from the difference between the inertial eigenvalues, 
$(I_j - I_k)$, as a proxy for the quadrupole moment.   Focusing on dark matter and gas just outside the virial radius of the galaxy, $(1<r/\Rvir<2)$, they found that the 
quadrupole moment is consistently higher for the cold gas than it is for the dark matter by a factor of $\sim1.5-2$.  Since these tidal torques may act prior to maximum, 
each stream may acquire a transverse velocity, so that it is no longer pointing directly at the center \citep[in agreement with previous work by][]{Pichon11}.
The highest angular momentum streams thus have spin parameters
as high as $\lambda\sim0.3$ upon crossing the virial radius, however, misalignment between multiple streams typically lowers the net spin parameter of all cold gas entering
the virial radius to a lower value of $\lambda\sim0.1$ \citep[in agreement with previous work by][]{Stewart13}.
\item While inflowing dark matter virializes once inside the virial radius, the cold streams penetrate the halo quickly, without becoming well--mixed with the pre-existing gas 
in the halo, resulting in a higher spin parameter for cold gas in the halo  $(0.1 < r/\Rvir < 1.0)$ by a factor of $\sim3$ when compared to the dark matter in the same volume 
(see Figure \ref{fig_gasspin}). The cold gas in the outer halo is also significantly more coherent than the dark matter, with a significantly smaller anti-rotating fraction.
\item The cold streams remain coherent, spiraling around the galaxy and sinking quickly towards the center of the halo.  As the streams blend and mix together, they often form into
what \cite{Danovich15} refers to as ``extended rings'' of inflowing cold gas, the radius of which is typically set by the pericenter of the stream contributing the most angular 
momentum.  These structures are typically warped with respect to the inner disk, in qualitative agreement with the ``cold flow disk'' structures reported 
previously by \cite{Stewart11b} and \cite{Stewart13}.  Similar structures have also been noted
as areas of interest in previous simulations: for example, the ``messy region'' of \cite{Ceverino10} or the ``AM sphere'' or \cite{Danovich12}.  Angular momentum in these 
extended rings is ultimately lost as a result of strong tidal torques from the inner disk on timescales of roughly one orbital time, allowing the extended disk to gradually align with 
the inner disk.
\item The angular momentum lost by the inspiraling cold gas can ultimately  be redistributed to both outflows and the dark matter.  The inner disk is subject to angular momentum 
redistribution and violent disk instabilities. 
\end{enumerate}

\cite{Teklu15} analyzed $\sim600$ intermediate resolution massive ($\Mvir>5\times 10^{10}\Msun$) halos from the \emph{Magneticum} simulation (Dolag et al. in preparation) over the redshift range $z=2-0.1$.
In agreement with previous results, they compared the spin parameter of all dark matter, stars, gas, cold gas, and hot gas in the virial radius (not cutting out the inner region of the halo where the galaxy resides), 
and found that the distribution of spins was well fit by lognormal distributions, with 
the gas (particularly the cold gas components) showing systematically higher spin than that of the dark matter, with the dark matter spin staying roughly constant with time, but the gas spin parameter
growing with time: $\lambda_{\rm cold} \sim2(3)\lambda_{\rm DM}$ at $z=2(0.1)$. They also noticed a dichotomy
in spin parameter with galaxy morphology: disk galaxies tend to populate halos with slightly higher spin parameters, and where there is better alignment between the angular momentum vector of 
the inner region of the dark matter halo versus that of the entire halo.

\begin{figure}[t]
\includegraphics[width=0.475\textwidth]{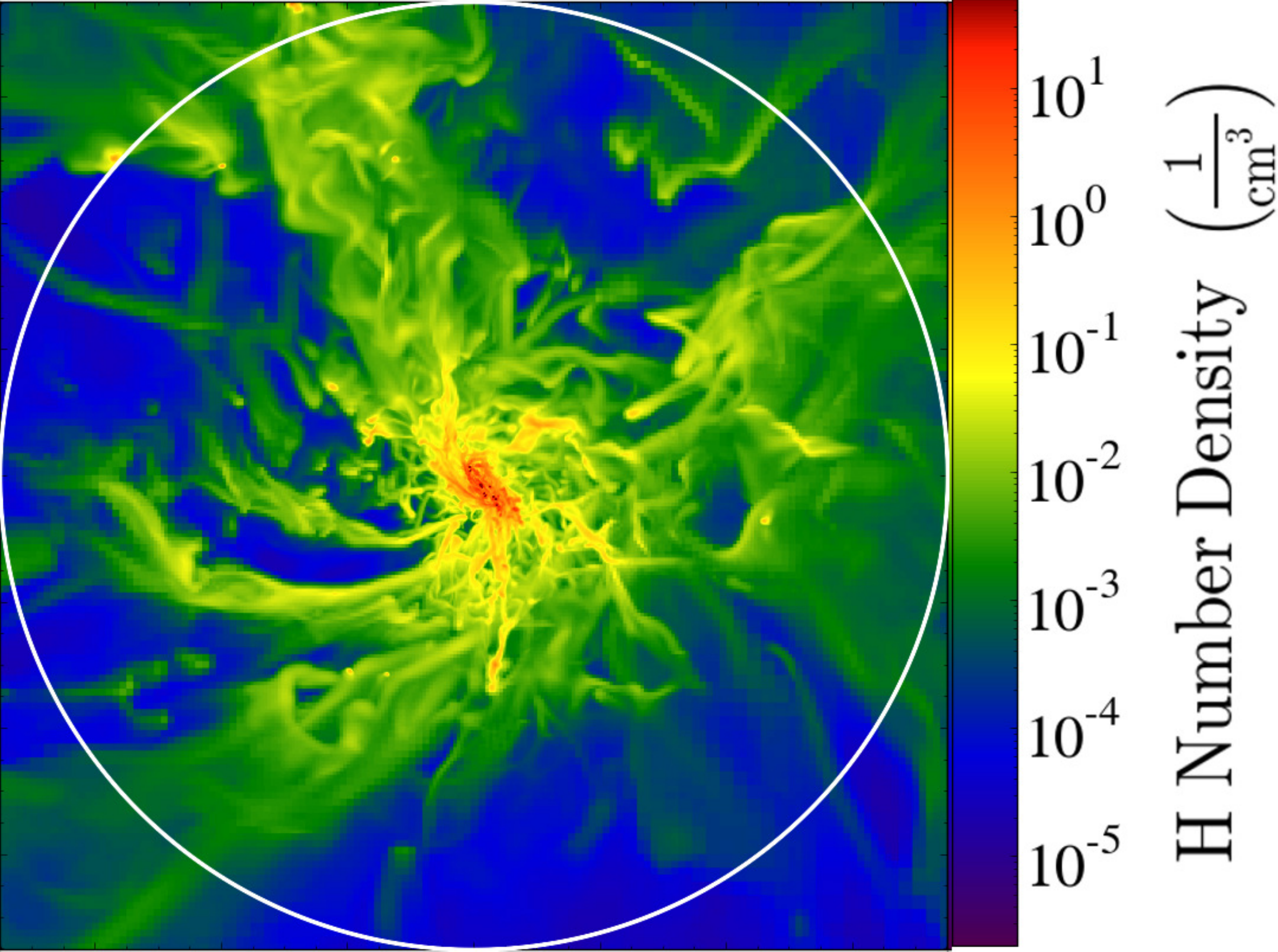}
\hspace{0.05\textwidth}
\includegraphics[width=0.475\textwidth]{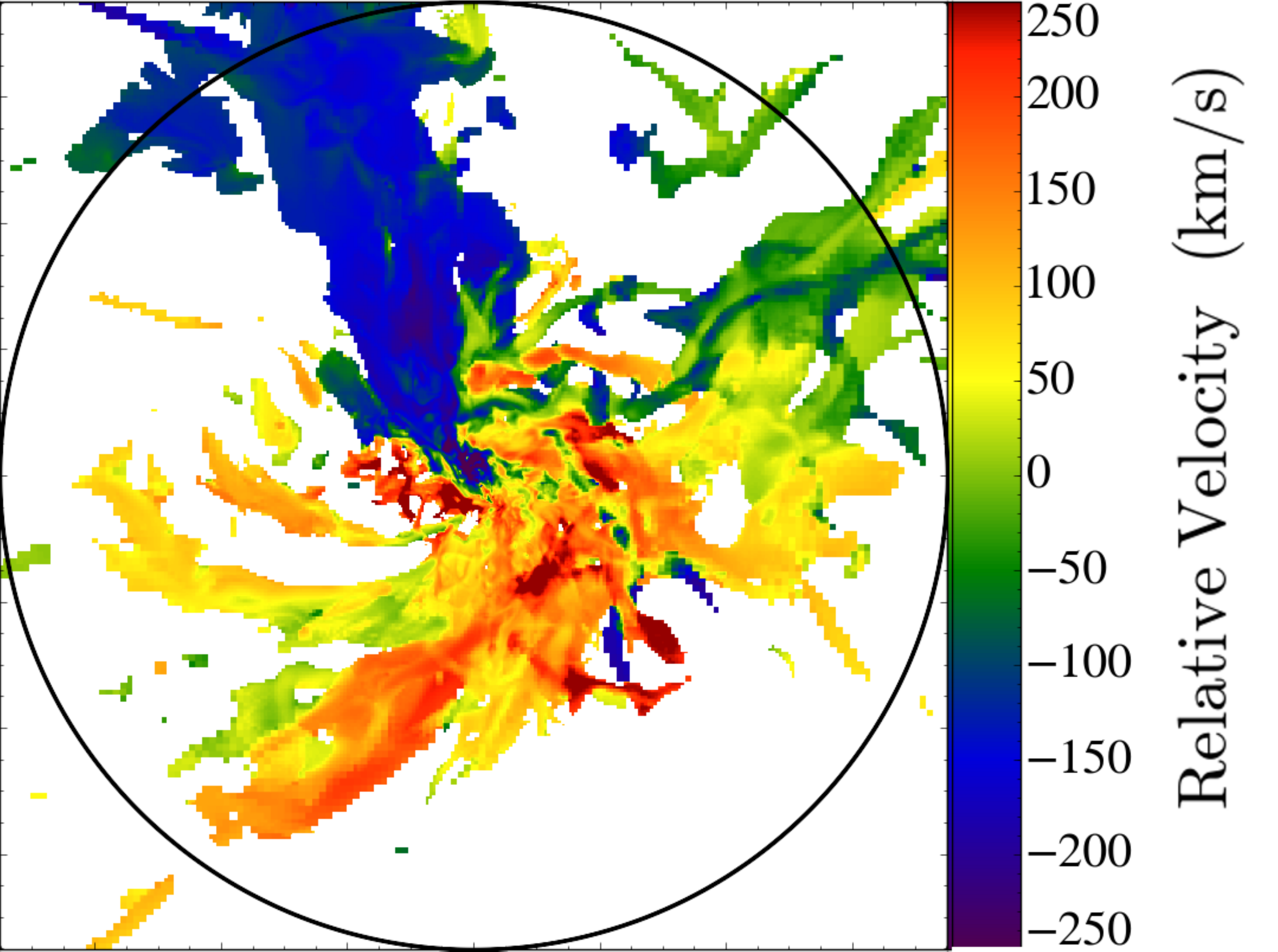}
\caption{Inspiraling cold streams in a galaxy simulation at $z=3$.  The halo virial radius is annotated by a circle in each panel.  Left panel shows projected 
H number density, and the right panel shows density--weighted line of sight velocity for gas with a minimum density threshold of $n_{\rm H}>3\times10^{-3}$ cm$^{-3}$
(approximately equivalent to $N_{\rm HI}\gtrsim10^{17}$ cm$^{-2}$).  The coherent bulk rotation of the inspiraling cold streams is apparent and should, in principle, represent 
an observable test of filamentary gas accretion in LCDM.}
\label{fig_coldflowdisks}       % Give a unique label
\end{figure}

In an effort to test whether this changing picture of angular momentum acquisition is sensitive to simulation code architectures or specific feedback implementations, 
\cite{Stewart16} carried out a code comparison of a single high resolution zoom--in simulation of a Milky Way sized halo 
(using common recent hydrodynamic/feedback implementations for each code, and utilizing identical analysis for each code) run with 
$\enzo$ \citep{Enzo}, $\art$ \citep{Art1,Art2}, $\ramses$ \citep{Ramses}, $\arepo$ \citep{Arepo}, and $\gizmo$ \citep{Hopkins15}.  
While many quantitative differences 
were apparent among the codes, agreements included the spin of cold halo gas being $\sim4$ times higher than the dark matter in the halo 
\citep[in agreement with previous work, e.g. Figure \ref{fig_gasspin}, taken from][]{Danovich15}, 
as well as the presence of inspiraling cold streams.  These inspiraling cold streams often form extended transient structures of high angular momentum cold gas, co-rotating with the galaxy along a preferred plane 
that is kinematically linked to inflow via large--scale cosmic filaments (see Figure \ref{fig_coldflowdisks} and discussion in \S\ref{coldflowdisks}). 
%structures with properties similar to previous work, 
The agreement among disparate simulation codes and physics implementations suggest that these aspects (at minimum) are likely to be robust predictions of 
galaxy formation in the Lambda Cold Dark Matter paradigm.

%>>>>>>>>>>>>>>>>>>>>>>>>>>>>>>>>>>>>>>>><<<<<<<<<<<<<<<<<<<<<<<<<<<<<<<<<
%>>>>>>>>>>>>>>>>>>>>>>>>>>>>>>>>>>>>>>>><<<<<<<<<<<<<<<<<<<<<<<<<<<<<<<<<

\subsubsection{Theoretical Predictions: High Angular Momentum, Co--rotation, and Inspiraling Cold Streams}
\label{coldflowdisks}
The most direct observable predictions of this new picture, specifically as it relates to angular momentum is thus not likely to come from studies of galaxies themselves 
(which represent a complex cumulative history of past angular momentum acquisition---including mergers, stream misalignments, etc.---as well as effects from stellar feedback and outflows),
but,  from observations of the circumgalactic medium.
In the canonical picture (outlined in \S\ref{galaxies}) cold gas in galaxy halos is thought to have cooled out of a virialized hot halo, and should have 
roughly the same angular momentum distribution as the dark matter.  In this new picture, cold gas in galaxy halos should have $\sim4$ times higher spin,
and often form coplanar structures of coherent inflowing gas, fueled by filamentary gas accretion.  
In an attempt to find a middle ground between the different terms in the literature for such structures, 
we will refer more generally here to these phenomena as resulting from inspiraling cold streams, since the degree to which 
these structures resemble the disk--like \citep{Stewart11b,Stewart13} or ring--like \citep{Danovich15} morphologies from previous work may be sensitive to specific hydrodynamic codes, 
feedback implementations, and possibly the halo mass scale involved.  

Figure \ref{fig_coldflowdisks} shows one example of inspiraling cold streams in a galaxy halos, taken from a cosmological hydrodynamic zoom--in simulation 
and visualized on the scale of the halo virial radius (denoted by the circle in each panel).  The left panel 
shows the projected H number density and the right panel showing the projected density--weighted 
line of sight velocity for all sight lines that meet a minimum column density 
threshold\footnote{In detail, a minimum 3D hydrogen density cutoff of $n_{\rm H}>3\times10^{-3}$ cm$^{-3}$ was implemented, however, this should correspond to a 
minimum hydrogen column density of $N_{\rm HI}\gtrsim10^{17}$ cm$^{-2}$ \citep[e.g.,][]{Altay11, Schaye01}} of $N_{\rm HI}\gtrsim10^{17}$ cm$^{-2}$.
The coherent rotational structure of the inspiraling cold streams (fueled and kinematically connected to the larger filamentary geometry) is apparent.  
Encouragingly, this type of extended rotational structure of inflowing gas bears a striking similarity to recent observations 
of giant protogalactic disks \citep[][]{Martin15,Martin16}, which have been detected in $\Mvir\sim5\times10^{12} \Msun$ halos at $z\sim2-3$.
In qualitative agreement with simulations, these observed disk--like structures extend to diameters 
of $\sim100$ kpc ($\sim\Rvir/2$), with rotational velocities of $\sim300$ km/s that show a kinematic connection to an inflowing filament, 
and have very high angular momentum (estimated $\lambda\sim0.1-0.3$), with orbital times comparable to the halo dynamical time.

In an effort to compare simulations to absorption line studies that match absorber kinematics to the rotation curve of the associated galaxy, 
\cite{Stewart11b} also created mock absorption sightlines to infer that, for inflowing gas, $\sim90\%$ of absorbers with $N_{\rm HI}\gtrsim10^{16}$ cm$^{-2}$ 
should have line of sight velocities completely offset from the system velocity of the galaxy in a single direction (per sightline) by $\sim100$ km/s,
with most of these absorbers roughly co--rotating with the galactic disk.  Again, the results from simulations are in encouraging agreement with 
recent absorption studies where the associated galaxy kinematics are known \citep[e.g.,][also see \S\ref{CGM_observations}]{Bouche16,Bowen16}, though
larger statistical samples of both observations and high resolution zoom--in simulations will be important for characterizing the level of agreement in detail.

\section{Summary and Conclusion}
\label{conclusion}
In this review of gas accretion and the angular momentum
of galaxies and galaxy halos, we began (\S\ref{DM}) by reviewing the origin of angular momentum in dark matter halos via Tidal Torque Theory, where large scale tidal torques before halo turnaround
set the initial angular momentum of a collapsing region based on the structure of large scale overdensities.  This ultimately sets a distribution of halo spin parameters in dark matter halos,
independent of halo mass, and with typical spins of $\lambda = j/\sqrt{2}\Vvir\Rvir \simeq 0.04$.

Under the canonical galaxy formation model (\S\ref{galaxies}), it is presumed that the angular momentum of inflowing gas matches that of the dark matter, shock--heats to the virial temperature of the 
halo, where the gas becomes well-mixed, before ultimately cooling out of the halo while conserving angular momentum to form a rotationally supported disk galaxy at the halo center.
Under this picture, the hot gas halo is expected to have roughly the same spin distribution of the dark matter ($\lambda_{\rm gas}=\lambda_{\rm DM}$), 
such that the disk galaxy that eventually forms should have a disk size 
of roughly $R_d\simeq\lambda\Rvir$.  While this picture is a good approximation for estimating galactic disk sizes in practice, there are growing number of 
observations (outlined in \S\ref{CGM_observations}) both in the local and distant universe that 
demonstrate the presence of significantly higher angular momentum material at large distances from the 
centers of galaxies, which might seem difficult to explain under the above scenario.

Our main emphasis in this review (\S\ref{CGM_theory}) has been a summation of recent findings from hydrodynamic cosmological simulations that 
suggest a modified picture for angular momentum 
acquisition, particular as it relates to predictions for the circumgalactic medium (CGM) of massive galaxies at $z\gtrsim1$.  This emerging picture, found in qualitative agreement among 
a variety of simulations---including large statistical samples of intermediate resolution simulations, smaller statistical samples of high resolution zoom-in simulations, and a wide 
range of code architectures and feedback physics implementations---is summarized as follows:
\begin{itemize}

\item The dense filamentary nature of cosmic gas accretion plays an important role in the properties of the CGM, with the geometry of filamentary gas responding differently to the initial tidal torques
that set angular momentum at early times (a factor of $\sim1.7$ enhancement versus the dark matter in the filament).  This angular momentum translates to a net transverse velocity of the filament and 
subsequently a non-zero impact parameter as the filamentary gas enters the virial radius of the halo.  

\item Material (both gas and dark matter) in a cosmic void that is initially farther away from the filaments gains a larger transverse velocity by the time it impacts the filament. 
As a result, the specific angular momentum of material freshly accreted to the virial radius (for both dark matter and gas) increases with cosmic time.  Compared to the cumulative 
average of the material that already exists in the halo, the spin of fresh accretion is also enhanced by a factor of $\sim2.5$.

\item Filamentary gas tends to have cooling times shorter than compression times for stable shocks at high-z, even for massive galaxies.  As a result, filamentary (``cold flow'') gas accretion 
sinks quickly (in $\le2$ times the freefall timescale) to the center of the halo upon entering the virial radius.  Thus, cold gas currently in the CGM of a given galaxy shows an enhanced
spin by a factor of $\sim4$ (compared to the dark matter).  This is a combination of the previous two effects mentioned above: 
1) the high intrinsic spin of cold filamentary gas and 
2) the enhanced spin for recent accretion to the halo (when compared to the dark matter halo, which probes a cumulative total of all past accretion).

\item One natural result of this high spin inflow with short sinking times (and thus, often a coherent direction for the angular momentum vector of all cold gas currently in the CGM), 
is that the inspiraling cold streams in the halo often form massive extended structures of roughly coplanar cold CGM gas, showing coherent rotation (see figure \ref{fig_coldflowdisks}) 
as the gas flows from the virial radius to the center of the halo.  While the existence of these structures seems robust in a qualitative sense, the exact nature and prevalence of these structures 
(and their implications for galaxy formation) are still unknown.  However, it is promising to note that recent observations of ``protogalactic disks'' seem to show qualitatively 
similar structures in the real universe.

\end{itemize}

It is important to note that much of emphasis in this modified picture for angular momentum acquisition 
has been on the process by which cold filamentary gas transitions from the cosmic web through the CGM on its way to the galaxy, particularly that this cold gas has significantly higher angular momentum 
while in the CGM than either the dark matter halo or the baryons in the galaxy.  
One important clarification is that this factor of $\sim4$ enhancement in cold halo gas spin versus dark matter is a result of the coupling of cold CGM gas properties 
(gas that is freshly accreted to the halo, and that is of a filamentary origin) working together to produce this enhancement. The same level of spin enhancement 
should \emph{not} hold for the cold gas in the \emph{galactic} region, as the baryons near the 
galactic center (or indeed within the galaxies themselves) are more likely to probe a prolonged accretion history from multiple ``modes'' of accretion,
and is thus more likely to mimic the spin of the dark matter halo, as expected from the canonical picture of galaxy formation.

For example, Figure \ref{fig_gasspin} showed the spin parameter distribution for 
cold halo gas, but did not include the inner region (i.e. $0.1 < r/\Rvir < 1$).
However, misalignment between the inspiraling cold streams and the baryons in the central region typically leads to significant vector cancellation, 
with a lower overall spin parameter for gas once the galactic region is included.  
Thus, while the mean spin parameter of cold halo gas shown in Figure \ref{fig_gasspin} is $\langle\lambda_{\rm cold}\rangle=0.11$, 
in a complementary panel in the same figure \citep[Figure 6 from][]{Danovich15} the mean spin parameter for all cold 
gas within the virial radius of the halo ($r<\Rvir$) is noticeably reduced: $\langle\lambda_{\rm cold}\rangle=0.086\sim2\langle\lambda_{\rm DM}\rangle$.
%, due in combination to 
%1) the significant misalignment between the cold halo gas and cold galactic gas and 
%2) the lower angular momentum content of the (cumulatively acquired) cold galactic gas versus the actively inflowing, recently accreted cold halo gas.  
Thus, in angular momentum studies where all material within the virial radius is included (including the galactic region), only this factor of $\sim2$ enhancement
of cold gas versus dark matter is expected.  For example, \cite{Teklu15} compared
of the spin of all cold gas within $r<\Rvir$ to that of the dark matter at $z=2$, finding that $\lambda_{\rm cold}=0.074\sim2\lambda_{\rm DM}$.
Similarly, \cite{ZjupaSpringel16} recently studied the angular momentum of dark matter halos 
and their baryons for $\sim320,000$ moderately high resolution halos
from the \emph{Illustris} simulation \citep{Illustris}---comparing 
all gas within the virial radius ($r<\Rvir$) and not making any distinction between cold 
versus hot gas components, finding that $\lambda_{\rm gas}\simeq0.1\sim2\lambda_{\rm DM}$, in agreement with other work reviewed here.

We also note that while the modifications suggested here for the standard picture of angular momentum acquisition in galaxy halos has strong implications for the angular momentum of baryons in the CGM,
it is unclear at this time how this modified picture directly impacts of the angular momentum of the galaxies that form at the center of the halo.  
If cold gas accretion onto galaxies typically has higher spin than the dark matter, but that angular momentum is subsequently 
lost from the galactic disk by strong torques from inspiraling cold streams, or redistribution of angular momentum via subsequent mergers, diffuse accretion and/or outflows, 
it may be that the similar spins for galactic disks and dark matter halos are merely the result of  
coincidence.  Alternatively, since the  galaxy that ultimately  forms at the center of a growing dark matter halo is the result of an extended, cumulative process, which must 
by its very nature account for misalignments in the  angular momentum direction of accretion over cosmic timescales, it may not be surprising (or coincidental) 
that the specific angular momentum of galactic disks are similar to their dark matter halos.  After all, the dark matter halo also probes the cumulative 
history of angular momentum acquisition over cosmic time.

%\bibliographystyle{apj}
%\bibliography{referenc-stewart}
%\input{referenc-stewart}

\end{document}